\documentclass[%
 reprint,
 amsmath,amssymb,
 aps,
nofootinbib]{revtex4-2}

\usepackage{graphicx}% Include figure files
\usepackage{dcolumn}% Align table columns on decimal point
\usepackage{bm}% bold math
\usepackage{color}
\usepackage{braket}
\usepackage{booktabs}
\usepackage{comment}
\usepackage{aas_macros}
\usepackage{url}
\usepackage{systeme}
\usepackage{gensymb}
\usepackage{CJK}

\begin{document}

%\preprint{APS/123-QED}

\title{Integral field spectroscopy with no IFUs: combining wide-field rotational\\ slitless spectroscopy with tomographic reconstruction}% Force line breaks with \\

\author{Francesco Sinigaglia $^{1,2,3,4}$}
\email{fsinigag@sissa.it}
\author{Jerry Jun-Yan Zhang $^{5,6}$}

\email{junyan.zhang@uwo.ca\\Both authors contributed equally to the work.}
\vspace{0.5cm}
\affiliation{
$^{1}$Institute for Fundamental Physics of the Universe (IFPU), Via Beirut 2, I-34151 Trieste, Italy}
\vspace{0.2cm}
\affiliation{
$^{2}$SISSA - International School for Advanced Studies, Via Bonomea 265, 34136 Trieste, Italy}
\vspace{0.2cm}
\affiliation{
$^{3}$INAF - Osservatorio Astronomico di Trieste, Via G. B. Tiepolo 11, I-34131 Trieste, Italy}
\vspace{0.2cm}
\affiliation{
$^{4}$INFN – National Institute for Nuclear Physics, Via Valerio 2, I-34127 Trieste, Italy}
\vspace{0.2cm}
\affiliation{
$^{5}$Department of Physics and Astronomy, The University of Western Ontario, 1151 Richmond St, London, ON N6A 3K7, Canada}
\vspace{0.2cm}
\affiliation{
$^{6}$Institute for Earth and Space Exploration, The University of Western Ontario, 1151 Richmond St, London, ON N6A 3K7, Canada}

\date{\today}% It is always \today, today,

\begin{abstract}
Among the astronomical spectroscopic techniques, Integral Field Spectroscopy (IFS) is regarded as one of the most versatile and powerful, but it is limited by small fields of view, complex instrument designs, and extremely high costs. We present \textsc{rossini}: the ROtational Slitless Spectrograph for INtegral field spectroscopy and Imager, a novel spectrograph design aimed at performing IFS without integral field units (IFUs) but in a slitless fashion instead. The device relies on generating an arbitrary number of independent detector images of the same field by rotating the whole telescope and/or the dispersion direction of the optical element (prism, grating, or grism) and on a postprocessing tomographic reconstruction algorithm yielding a full IFS datacube, with arbitrary angular and spectral resolution defined by the user. By combining the very high efficiency of slitless spectroscopy with the high information content of IFS, \textsc{rossini} is the ideal instrument to perform IFS in wide fields. We first develop the mathematical formulation of the problem and derive the solution as a linear matrix inversion. Then, we provide an interpretation of \textsc{rossini} as a particular application of tomography and leverage this to propose a practical numerical solution based on iterative reconstruction. We test this novel conception through a series of numerical experiments: we first generate toy datacubes on the sky, then simulate the spectrograph and the corresponding detector images, and finally apply tomographic reconstruction, trying to recover the input datacube. Conceptually, the rotational slitless spectrograph can handle datacubes with an arbitrary pixelization. The number of needed rotations can be easily computed from the relative pixelization of the datacube and the pixel number of the detector. From the numerical experiments, \textsc{rossini} turns out to be able to reconstruct the benchmark datacubes with percent accuracy and in just a few hundred iterations, using negligible computational resources. We conclude that this novel conception is a promising way forward for future spectroscopic facilities, as it allows us to perform wide-field IFS in an efficient and cheap fashion by relying mostly on existing slitless spectrograph technologies.

\end{abstract}

\maketitle

%\tableofcontents
% *********************************
% *********************************
% *********************************
\section{Introduction}

Integral field spectroscopy (IFS) is a cornerstone technology of modern astrophysics, especially in studying extended objects, such as nebulae, galaxies, or a crowded cluster of stars or galaxies in one shot. IFS produces a $3$D datacube including both the $2$D spatial information and $1$D spectral information. 

Current IFS relies on integral field units (IFUs) which can have one of the following designs: a microlens array, a fibre bundle, or a mirror slicer. The first IFU idea came from G. Courtes in 1982, using a lenslet array and had been realised in Traitement Intégral des Galaxies par l'Etude de leurs Raies \citep[TIGERS; ][]{bacon1995tiger}  on the 3.58-m Canadian-France-Hawaii Telescope. Both the microlens array and fibre bundle share the same logic: dispersing light from a finite number of large $1$D spatial elements onto a $2$D detector. 

The Multi Unit Spectroscopic Explorer \citep[MUSE;][]{bacon2010muse} and the K-band Multi Object Spectrograph \citep[KMOS;][]{Sharples2013} on the Very Large Telescope, instead, adopt mirror slicers, which divide the field of view (FoV) into a series of narrow $1$D slices and project them onto $2$D detectors for dispersion. However, the $3$D datacube inevitably comes at the cost of either lower spatial resolution (larger pixels) or smaller FoV. This is the fact limited by the detector dimension. In addition, these traditional IFUs are usually expensive and technically demanding to manufacture. To have a relatively bigger FoV, MUSE and KMOS even equipped 24 mirror slicer IFUs combined multiple detectors, resulting in high cost and system complexity. Scaling up these IFU systems to even wider fields would make it technically challenging and very costly.

On the other hand, slitless spectroscopy has been widely employed in the last decades. Thanks to its cost-efficiency excellent trade-off, its optical simplicity which maintains the light throughput high, and its ability to survey wide field, it has been the obvious choice to deploy survey instruments on the main space telescopes. Slitless spectroscopy, however, does not perform well in crowded fields, as the spectra overlap on the detector. To alleviate this problem, a number of techniques have been proposed to perform partial spectra decontamination. In space telescopes, it can be achieved by either changing the grism and the dispersion orientation with a wheel, or by performing a roll movement of the whole telescope. For instance, the Near-Infrared Spectrometer and Photometer \citep[NISP,][]{euclid2025nisp} mounted on {\it Euclid} is equipped with a wheel carrying one blue grisms and three red grisms. The Near-Infrared Imager and Slitless Spectrograph \citep[NIRISS,][]{Doyon2023} mounted on the James Webb Space Telescope (\textit{JWST}) have two grisms rotated by $90^\circ$. Rotation around the roll angle are instead possible and used for spectra decontamination for all the main space telescopes:  the Hubble Space Telescope \citep[e.g.,][]{Kummel2009,Outini2020}, the  GALaxy Evolution Explorer (GALEX\footnote{\url{https://www.galex.caltech.edu}}), {\it JWST}\footnote{\url{https://jwst-docs.stsci.edu/methods-and-roadmaps/jwst-wide-field-slitless-spectroscopy}}, {\it Euclid} \citep{Outini2020}, and Roman \citep{Outini2020}. This concept can be translated to ground-based telescopes into a rotation of the whole telescope through the field rotator. However, these cases relying on very few rotations can rarely perform a full decontamination, and their usage to perform IFS has never been explored yet.

In this paper, we seek to exploit the rotational working principle of slitless spectrographs to perform IFS. In particular, we propose to leverage the fact that dispersing the light in different directions through a rotation of the optical dispersion element and/or the whole telescope produces independent detector images. We show that a sufficient number of detector images at different rotation angles provides enough linearly-independent constraints to be able to write a linear system of equations that mathematically expresses the flux in a datacube with arbitrary pixelization defined by the user as a function of the observed detector images. In this sense, it effectively represents an IFS technology with a wide-field design coming from slitless spectroscopy. We also show that this problem consists in a tomographic reconstruction case, thereby offering the possibility of exploiting the numerical techniques usually adopted in tomography to reconstruct the datacube. 
%We also show that, because the reconstruction to build the datacube consists of an operation performed a posteriori, it allows to set an arbitrary angular and spectral resolution, enabling having adaptative pixeling with non-trivial geometries. This fact is especially important because the pixel resolution of the datacube dictates the number of needed rotation, allowing to find the optimal trade-off between quality and cost.

The paper is organized as follows. In Section \ref{sec:concept} we outline the novel conceptual description of the instrument. In Section \ref{sec:math} we develop the mathematical formalism underlying this work. Section \ref{sec:methods} presents a proof of concept of the instrument and Section   \ref{sec:discussion} the discussion of the results. We conclude in Section \ref{sec:conclusion}.

%--------------------------------------------------------------------

\section{Conceptual description}
\label{sec:concept}

\begin{figure*}
    \centering
    \includegraphics[width=0.49\linewidth]{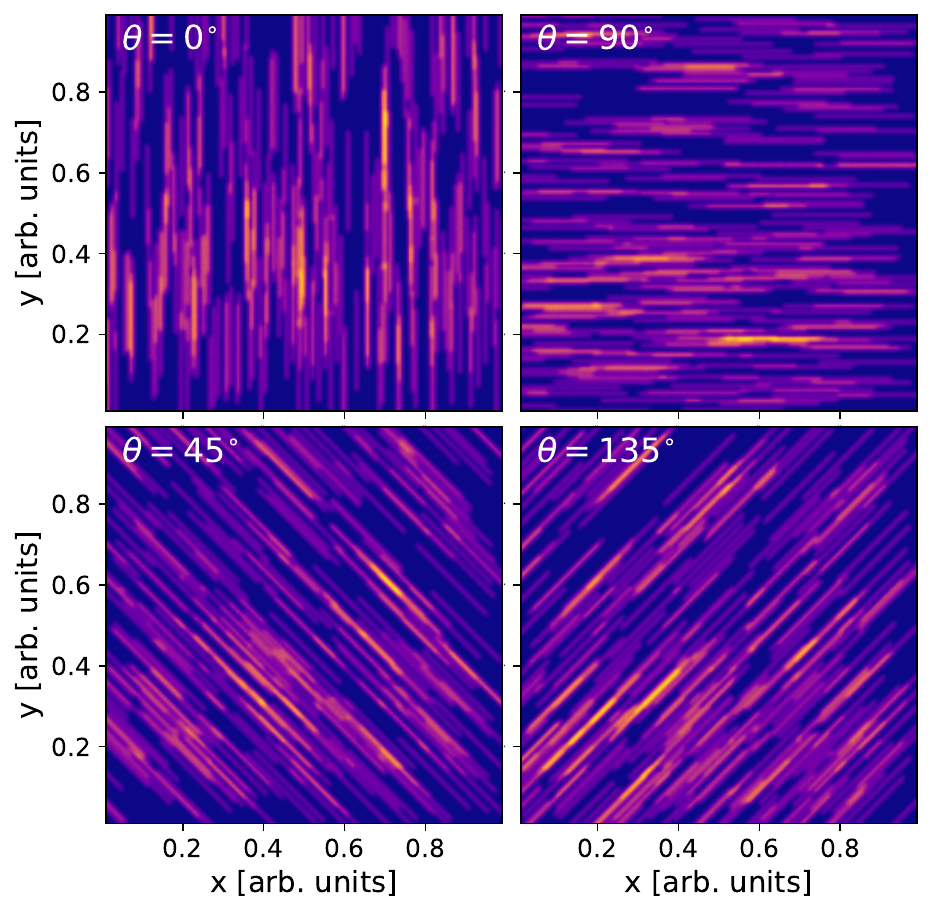}
    \includegraphics[width=0.49\linewidth]{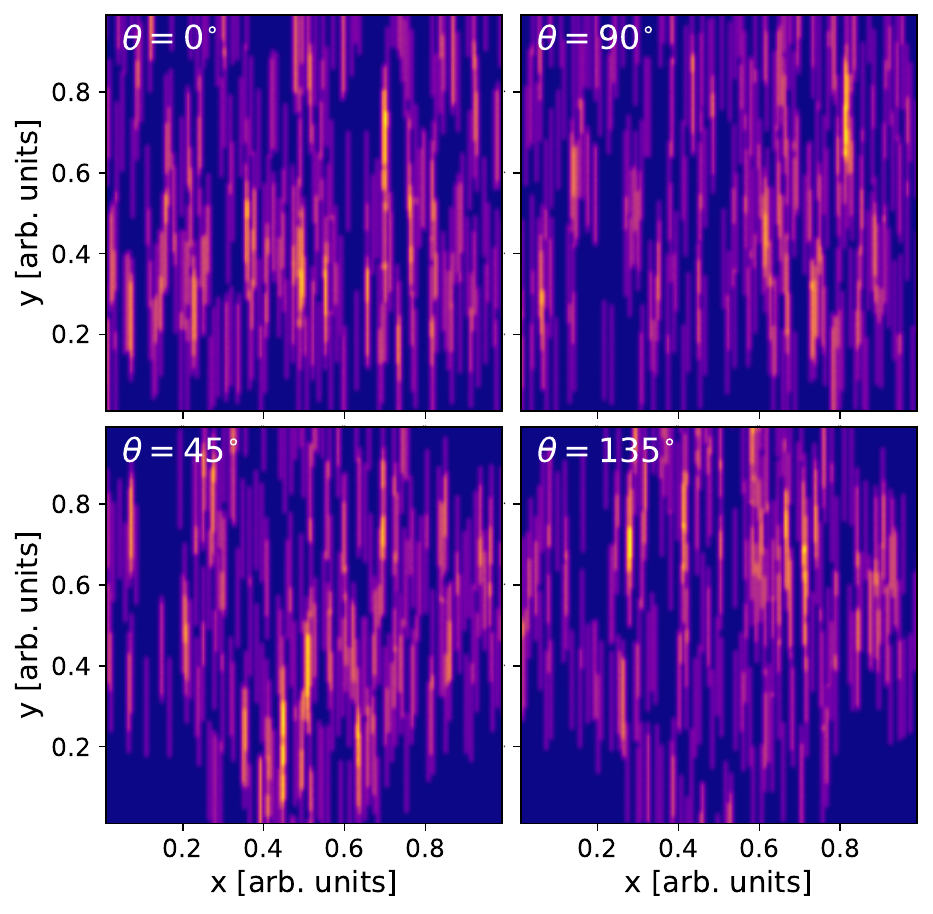}
    \caption{Left: On-detector images of a field with $300$ stars, in the case for which the dispersion element is rotated with respect to the detector, for four different rotation angles.%: $\theta=0^\circ$ (top left), $\theta=45^\circ$ (bottom left), $\theta=90^\circ$ (top right), and $\theta=135^\circ$ (bottom right). 
    Right: the same but in the case for which the aperture is rotated with respect to the field.%, for four rotation angles: $\theta=0^\circ$ (top left), $\theta=45^\circ$ (bottom left), $\theta=90^\circ$ (top right), and $\theta=135^\circ$ (bottom right).
    }
    \label{fig:ccd_sims_rot}
\end{figure*}

% \begin{figure*}
%     \centering
%     \includegraphics[width=0.7\linewidth]{ccd_simulations_fieldrot.pdf}
%     \caption{On-detector images of a field with $300$ stars, in the case for which the aperture is rotated with respect to the field, for four rotation angles: $\theta=0^\circ$ (top left), $\theta=45^\circ$ (bottom left), $\theta=90^\circ$ (top right), and $\theta=135^\circ$ (bottom right).}
%     \label{fig:ccd_sims_rotfield}
% \end{figure*}

In this section, we provide a conceptual description of the novel spectrograph design---\textsc{rossini}: the ROtational Slitless Spectrograph for INtegral field spectroscopy and Imager---presented herein. As anticipated, the idea behind this work is to take advantage of existing simple slitless spectroscopy technology to reconstruct a complete spectral datacube $({\rm RA},{\rm DEC}, \lambda)$ in an IFS fashion without explicitly employing traditional IFUs. By rotating the spectrograph with respect to the detector and/or by rotating the aperture with respect to the sky, one can obtain independent on-detector images. In particular: 
\begin{itemize}
    \item by rotating the spectrograph with respect to the detector, the incoming light on the spectrograph is dispersed onto the detector in different directions depending on the relative angle between the detector and the light dispersion direction of the optical dispersion element. This case is illustrated in the left panel of Figure~\ref{fig:ccd_sims_rot}, for four different rotation angles; 
    \item by rotating the whole aperture, the center of the field stays fixed and the celestial sources have their light focused on different positions on the spectrograph. The spectrograph maintain the same orientation in the detector frame. This case is illustrated in the right panel of Figure~\ref{fig:ccd_sims_rot}.
\end{itemize}
 In this way, even in a situation where the spectra from two sources are superposed on the detector, deploying several angles allows disentangling the two lights from the sources and enabling the full reconstruction of the two spectra. This argument can be generalized for an arbitrary number of sources, provided that a suitable number of rotations of the spectrograph with respect to the detector and/or of the telescope are performed. %In the science case that we present in this work, we consider a fictitious datacube on the detector plane.%, in which each pixel corresponds to a source and its spectrum gets spread onto the detector. Even though the image is not real but just an arbitrary representation, it effectively allows us to perform IFS if we are able to fully reconstruct the datacube.

%\JZN{ suggests that in Fig 1 and 2 we also put a source map (1 map for scenario 1 and 4 maps for senario 2)}\FSN{We could do it}

Each rotation the spectrograph is deployed, the spectra coming from different %`pixels' 
positions on the spectrograph plane will %contribute to different %pixels
generate mutually independent spectra on the detector. In this way, each rotation of the spectrograph provides a new set of constraints, where the flux received by each pixel can be expressed as a linear combination of the spectra from different pixels on the spectrograph plane. From this perspective, the problem reduces to determining the minimum number of rotations needed to %fully 
reconstruct the datacube, i.e. providing enough spatial and spectral resolutions. While we refer the reader to Section~\ref{sec:math} for a detailed mathematical description of the problem, we anticipate from now that this translates into determining the minimum number of rotations to obtain enough linearly independent equations to fully solve the problem.   

We highlight from now that this novel spectrograph conception offers the following advantages over traditional IFUs, which would imply a fixed pixelization in the FoV and spectral resolution:
\begin{itemize}
    \item because the datacube that we seek to reconstruct is not physical but just a representation of the continuous 2D distribution of incoming light on the detector, this allows us to choose every time the most suitable pixelization of the datacube depending on the specific observational target. This feature is quite attractive, since for low-resolution observations it implies that a lower number of rotations will be needed;
    \item in connection with the previous point, the datacube can have an arbitrary pixelization, and in particular it is not bound to have a regular rectangular gridding. This offers the possibility of employing an adaptive pixelization of the 3D datacube, both on the sky plane and in the spectral direction, observing at low resolution regions of lower interest and at high resolution others of higher relevance. This is the case e.g. when a portion of the image on the sky features pronounced clustering properties. In such a case, high resolution can be adopted in the most clustered regions --- where higher resolution is typically desired --- without the need of having the same high resolution everywhere in the image. Along the spectral axis, this property can be exploited to better resolve particular features such as e.g. emission and absorption lines or the continuum breaks (e.g. Balmer break), while the resolution can be relaxed in slowly-varying spectral intervals of the continuum;
    \item as will be shown mathematically in Section~\ref{sec:math}, because the image on the spectrograph plane is fictitious, we can achieve a sub-pixel resolution, overcoming the limits implied by the detector pixel size. % and the actual PSFs. \JZN{  our approach can go beyond PSF I think.} \FSN{It can go beyond the PFS, meaning that you can have an arbitrarily small pixel and "oversample" the PFS. But the PFS is dictated by $\sim \lambda/D$, so by the aperture of the telescope. In this sense, we cant do much.} 
    This is possible thanks to the arbitrarily high number of independent constraints that we can get from different rotations.  
\end{itemize} 

%\FSN{Provide here the conceptual description to provide an intuition of what we want to do and why. Then in the next section we do the Math.}

%\FS{
%I think we have to highlight the following important points:
%\begin{itemize}
%    \item we do IFU with slitless spectroscopy, which means that we  do not need fibers;
%    \item the idea is to reconstruct the datacube on the spectrograph plane which actually doesn't exist, but arises once we properly discretize the incoming light
%    \item given that the datacube is fictitious, we can choose an arbitrary pixelization both in the sky plane and in the spectral direction. This gives a lot of freedom because: (i) it is not bound to the detector specs, (ii) it allows to easily tune the exposure time (i.e. the number of rotations) and reduce the complexity by just matching the desired spatial and spectral resolution;
%    \item the datacube can have higher spatial resolution than the detector (magic!)
%    \item once can perform an adaptative pixelization of the sky plane to account for local density variations (important when clustering measurement are involved, mention fiber collision problem for cosmological surveys here). 
%\end{itemize}
%}

% -------------------------------------------------------------------

\section{Mathematical description}
\label{sec:math}

In this section, we provide a mathematical formulation of the problem that we wish to solve by means of the proposed rotational spectroscopy.

\subsection{Rotational slitless spectroscopy as an inversion problem}\label{sec:inversion}

%\begin{figure}
%    \centering
    %\includegraphics[width=\linewidth]{illu.png}
    %\caption{Illustration of an simple case, with three light beams of a continuum, an emission, an absorption spectra, being dispersed by two different angles between the disperser plane ($u$, $v$) and the detector plane ($x$, $y$). The physical disperser is hidden. The two overlapped emission and absorption spectra in the left were resolved in the right after the disperser rotated.}
    %\label{fig:placeholder}
%\end{figure}

Let us $(u,v)$ and $(x,y)$ be the coordinates describing the focal plane before the spectrograph and the detector plane, respectively, and $\lambda$ be the wavelength, i.e. the coordinate along the spectral axis. Let us $\mathcal{G}(u,v,\lambda;x,y), ~\mathcal{G}:(u,v,\lambda)\rightarrow (x,y)$ ---called the `response function' --- be the mapping between the focal plane and the detector coordinates. The spread function $\mathcal{G}$ will map the incoming light on the focal plane onto the detector through a dispersion element, in a way which depends on the spectrograph specific features and configuration. In particular, $\mathcal{G}$ will depend on the dispersion power of the spectrograph and on the angle $\theta$ between the dispersion element axis and the $(u,v)$ plane, which will change when the spectrograph is rotated with respect to the detector, as proposed in this work. In the detector, the flux $f$ observed at coordinates $(x,y)$ is given by the following expression:
\begin{equation}\label{eq:ccd_flux}
    f(x,y) = \int d\lambda\int du \int dv ~ \mathcal{G}(u,v,\lambda;x,y)~f'(u,v,\lambda) \quad, 
\end{equation}
ignoring the throughput of the system including the filter transmission, grism efficiency, detector response etc., which can be calibrated afterwards. 
This equation expresses the flux at point $(x,y)$ on the image plane as the continuous sum of the flux $f'(u,v,\lambda)$ at all the points $(u,v,\lambda)$ whose light is conveyed to $(x,y)$ by the dispersion element.

In the discrete limit, corresponding to an arbitrary pixelization on both the spectrograph and detector planes, the equation above reduces to:
\begin{equation}
    f(x,y) = \sum_i\sum_j\sum_k \mathcal{G}(u_i,v_j,\lambda_k;x,y)~f'(u_i,v_j,\lambda_k) \quad ,
\end{equation}
i.e. the integral reduces to a sum over a finite number of points. 
The goal of our treatment is to fully determine the spectrum of the incoming light at coordinates $(u_i,v_j,\lambda)$ while in principle the incoming light is a continuous 2D spatial distribution. For simplicity, let us assume that these points are given by a regular and orthogonal pixelization with $U$ and $V$ grid points along the $u$ and $v$ directions, respectively.  %While the incoming light on spectrograph plane comes in principle in a continuous 2D distribution, we need to discretize it in some way in order to be able to solve the problem and perform IFU with this novel rotational spectroscopy. 
We assume the $U \times V$ rectangular gridding described above to mimic a realistic case, but we stress that discretizing the spectrograph plane with a pixelization of arbitrary geometry would be achievable without loss of generality by just appropriately changing the functional form of the mapping $\mathcal{G}$ from the spectrograph to the detector plane.   

Let us also discretize each spectrum into $l$ wavelength bins, i.e., an arbitrary spectral resolution. Then we are interested in determining $U\times V\times l$ unknowns. In order to solve the system, we must then have a set of $U\times V\times l$ linearly-independent equations. 

Let us assume that the detector is composed of $X\times Y$ pixels in the $x$ and $y$ directions, respectively. Then, each exposure provides us with $X\times Y$ independent equations. At this point, the number of times $n$ that we need to rotate the spectrum is automatically determined by the following equation:
\begin{equation}
n \times X \times Y \geq U\times V \times l \quad .
\end{equation}

Solving for the minimum $n_\mathrm{min}$ under an ideal condition:
\begin{equation}\label{eq:num_rot}
    n_\mathrm{min} = \frac{U\times V\times l}{X\times Y} \quad , 
\end{equation}
where this number needs to be rounded to the closest higher integer. This result, that we have derived heuristically based on intuitive arguments, mathematically corresponds to the Hilbert projection theorem applied to linear inversion problems.

As anticipated, once $U\times V\times l$ equations and the function $\mathcal{G}$ are specified,  the linear system of equations can be readily solved:

{\footnotesize
\[
\systeme{f{(x_0,y_0,\theta_0)}=\sum_{i=1}^{U} \sum_{j=1}^{V}\sum_{k=1}^{l} \mathcal{G}_{\theta_0}{(u_i,v_j,\lambda_k;x_0,y_0)}~f'{(u_i,v_j,\lambda_k)}, f{(x_0,y_0,\theta_1)}=\sum_{i=1}^{U} \sum_{j=1}^{V}\sum_{k=1}^{l} \mathcal{G}_{\theta_1}{(u_i,v_j,\lambda_k;x_0,y_0)}~f'{(u_i,v_j,\lambda_k)},\dots, f{(x_0,y_0,\theta_{n})}=\sum_{i=1}^{U} \sum_{j=1}^{V}\sum_{k=1}^{l} \mathcal{G}_{\theta_{n}}{(u_i,v_j,\lambda_k;x_0,y_0)}~f'{(u_i,v_j,\lambda_k)},
f{(x_0,y_1,\theta_0)}=\sum_{i=1}^{U} \sum_{j=1}^{V}\sum_{k=1}^{l} \mathcal{G}_{\theta_0}{(u_i,v_j,\lambda_k;x_0,y_1)} ~ f'{(u_i,v_j,\lambda_k)},
\dots,
f{(x_0,y_1,\theta_{n})}=\sum_{i=1}^{U} \sum_{j=1}^{V}\sum_{k=1}^{l} \mathcal{G}_{\theta_{n}}{(u_i,v_j,\lambda_k;x_0,y_1)}~f'{(u_i,v_j,\lambda_k)},
\dots,
f{(x_X,y_Y,\theta_{0}})=\sum_{i=1}^{U} \sum_{j=1}^{V}\sum_{k=1}^{l} \mathcal{G}_{\theta_{0}}{(u_i,v_j,\lambda_k;x_X,y_Y)}~f'{(u_i,v_j,\lambda_k)},
\dots,
f{(x_X,y_Y,\theta_{n}})=\sum_{i=1}^{U} \sum_{j=1}^{V}\sum_{k=1}^{l} \mathcal{G}_{\theta_{n}}{(u_i,v_j,\lambda_k;x_X,y_Y)}~f'{(u_i,v_j,\lambda_k)}
}
\]
\label{eq:main_system}
}

Solving for $f'(u_i,v_j,\lambda_k) ~\forall ~ i,j,k$ fully specifies the datacube that we are interested in.

Thus far we have been treating the system by neglecting %the effects of the point spread function $\mathcal{P}$ (hereafter PSF) and of
the noise. %The effect of the PSF can be mathematically viewed as an additional convolution that acts on the spectrograph plane and distorts the photon paths towards the detector. For this reason, while it spreads flux coming from different points $(u,v)$ on the detector plane, on the spectrograph plane it can be reabsorbed in $\mathcal{G}$ as: $\mathcal{G}'=\mathcal{P}\circ \mathcal{G}$. At the level of the solution of the system of equation above, the PSF changes only the coefficients $\mathcal{G}_{\theta_0},u_i,v_j,\lambda_k;x_0,y_0)$, but it does not alter the linearity of the system. Therefore, the PSF is automatically accounted for once it has been included in the spread functions $\mathcal{G}$.
The noise acts instead on the detector plane, i.e. on the left-hand side of the equations above. In this sense, we can express the flux at point $(x,y)$ and given a rotation angle $\theta$ of the spectrograph as $f(x,y,\theta)=\tilde{f}(x,y,\theta)+\epsilon(x,y,\theta)$, where $\tilde{f}(x,y,\theta)$ and $\epsilon(x,y,\theta)$ are the true `noiseless' flux  and the noise at pixel $(x,y)$, respectively. $\epsilon(x,y,\theta)$ includes the photon Poisson shot noise $\sqrt{\tilde{f}(x,y,\theta)}$ and detector noises, in general. This implies that, because we observe $f(x,y,\theta)$ and hence cannot separate  $\tilde{f}(x,y,\theta)$ and $\epsilon(x,y,\theta)$, the noise contribution will be mapped onto $f'(x,y,\theta)$ through the solution of the system of equations above. In high signal-to-noise ratio (hereafter S/N) conditions $\tilde{f}(x,y,\theta)\gg \epsilon(x,y,\theta)$ --- which is the case of e.g. very bright sources --- the noise contribution can be neglected and it does not represent an issue. Although in low S/N cases, the noise contributes with a non-negligible uncertainty on $f(x,y,\theta)$, which is non-linearly mapped onto $f'(u,v,\lambda)$, we can approximately solve the equation. While we leave the detailed study of the impact of noise for future work and focus in this paper on an ideal high-S/N scenario, we caution that this may represent a challenge. In this sense, we have worked out the solution of the problem adopting an inversion strategy. Nonetheless, the noise can be properly modelled in forward-modelling strategies seeking to reconstruct the datacube $(u,v,\lambda)$ in a Bayesian fashion, by adopting gradient-based methods such as Hamiltonian Monte-Carlo sampling.

\subsection{Connection to tomographic reconstruction}\label{sec:tomography}
%\FSN{New subsection added, very important as it provide a natural mathematical background to show that not only is our problem well-defined, but the existence of the solution also have an elegant mathematical proof, as well as an efficient computational implementation.}

From a different perspective, the proposed rotational slitless spectroscopy is mathematically exactly equivalent to a $3$D Radon transform. Briefly, the Radon transform in an integral transform which maps a function $f(x,y,z)$ defined in a $3$D space onto a function $\mathcal{R}[f]$ which is equal to the line integral of the function over one line in the $2$D plane:
\begin{equation}
    \mathcal{R}[f](p, \vec{n}) \int_{\vec{x}\cdot\vec{n}=p} f(\vec{x})~{\rm d}S \quad ,
\end{equation}
where $\vec{x}$ denotes an arbitrary point in the $3$D space, $\vec{n}$ the normal vector to the plane along which the integration is performed, and $p$ the offset along the normal direction. In this sense, the flux measured on the detector at a given point $(x,y)$ expressed in Eq.~\ref{eq:ccd_flux} can be interpreted as integral along a plane in $(u,v,\lambda)$ space, determined by the grism dispersion and the rotation angle $\theta$, where different rotations correspond to different normal vectors $\vec{n}$ of planes in the 3D Radon transform and therefore rotating the instrument/dispersion axis means rotating the normal vector.   

This interpretation allows us to straightforwardly connect the novel rotational slitless spectroscopy presented herein to the well-known field of tomography, both at conceptual and at a methodological level. Tomographic reconstruction --- widely used in the medical imaging computed tomography (CT hereafter) \citep[see e.g.,][for a review]{Assili2018} --- consists in an ensemble of techniques which aim at reconstructing a $3$D object from a finite series of $2$D projections. 

The connection between rotational slitless spectroscopy and tomography offers three following important advantages:
\begin{itemize}
    \item it provides a solid mathematical backbone supporting the robustness of this spectroscopy techniques as a valid procedure to reconstruct a full datacube in IFS fashion;
    \item interpreting an on-detector image obtained at given rotation angle $\theta$ in a Radon-transform fashion, i.e. as a particular case of a family of $2$D projections of the full $3$D datacube, the Fourier transform of each on-detector image gives a slice of the 3D Fourier transform of the datacube and different angles $\theta$ sweep different planes in Fourier space. In this sense, the best choice for the angle values $\theta$ corresponds to the one that provides the most complete coverage in Fourier space, which naturally corresponds to uniform sampling in the interval $[0,\pi)$. Failing to properly sampling the Fourier space leads to the so-called `missing-cone tomography'; 
    \item instead of solving our problem directly as described in Section~\ref{sec:inversion} via linear inversion by building the full response matrix, whose dimension grows very quickly with the size of the datacube, we can leverage a variety of efficient computational and methodological techniques usually adopted in tomography to solve our problem. We discuss this point more in detail in Section~\ref{sec:methods}.  
\end{itemize}

% *****************************
% *****************************

%\subsection{Noise model}\label{sec:noise}

%--------------------------------------------------------------------

\section{Methods and validation}\label{sec:methods}

In this section, we describe the methodology that we adopt herein to simulate and validate the novel spectrograph concept.

\subsection{Spectrograph dispersion}\label{sec:spec_dispersion}

%\JZN{  Is this only for scenario 1 ?} \FSN{It's for both. The first set of equations are for the rotation of the optical element, the second set of equations are for the case in which the field is rotated.}

We model the spectrograph dispersion as follows. We neglect here the PFS and other possible distortions, in such a way that $\mathcal{G}$ effectively corresponds to only the dispersion. %Given a photon of coordinates $(u,v,\lambda)$ on the sky and $(x,y)$ the CCD coordinates of the photons after having been dispersed by the optical element, the operator that maps onto the detector is $\mathcal{G}:(u,v,\lambda)\rightarrow(x,y,)$. 
%For simplicity, we assume that the spectra rotate about the projected position of its pivot wavelength $\lambda_c$ on the detector.
In particular, $\mathcal{G}$ for the rotation of the dispersion element is defined as: 
\begin{align*}
    x &= u + d\cos(\theta) \\ y &= v + d\sin(\theta)
\end{align*}
while for the rotation of the whole telescope is:
\begin{align*}
    x &= \bar{x} + (u-\bar{u})\cos(\theta) -(v-\bar{v}) \sin(\theta) + d\\ y &= \bar{y} + (u-\bar{u})\sin(\theta) +(v-\bar{v}) \cos(\theta)
\end{align*}
where $\bar{x}$ and $\bar{y}$ denote the coordinates of the center of the detector, $\bar{u}$ and $\bar{v}$ the coordinates of the center of the field, $\theta$ is the rotation angle  and $d=\sqrt{(x-u)^2+(y-v)^2}$ is the dispersion and is assumed to be linearly proportional to the wavelength as $d=k(\lambda-\lambda_c)$, with $k$ a constant specific to the considered instrument and $\lambda_c$ is the pivot wavelength and depends again on the design and material of the dispersion element. We notice that in this ideal spectrograph design with no distortions nor aberrations, $\lambda_c$ corresponds to a pivot wavelength at which the incoming light is perpendicular to the focal plane onto the detector, preserving the same detector coordinates, independently of the position angle of the dispersion element. Therefore, in the first scenario described above, each spectrum will rotate around the detector point defined by $\lambda_c$. The constant $k$ describes the resolution power, with the idea that one would like the largest possible dispersion (corresponding to the largest possible resolution), without having the issue that spectra were projected outside of the detector. Here, we heuristically assume $k=\max(x,y)/[4\times(\lambda_{\rm max} - \lambda_{\rm min})]$ as an optimal trade-off in the spreading power of the spectrograph. Normally, this implies a low resolution power. However, this is just an arbitrary choice, and we have explicitly verified that the results presented in Section~\ref{sec:exp} depend only very weakly on this choice through the steepness of the gradients, which is a purely numerical problem that can be solved in other ways.

Therefore, to forward-model the problem from the sky to the detector --- both to generate the detector simulations related to the numerical experiments (see  Section~\ref{sec:exp}) and to compute the forward operator in the tomographic reconstruction (see Section~\ref{sec:tomography_solution}), we first initialize a collection of on-sky photons according to a given source and then spread them onto the detector by applying the equations above.  

\subsection{Discretization of the response function}\label{sec:spec_spread_function}

A key item of the linear system formalized above consists in discretizing the response function $\mathcal{G}^\prime_\theta(u,v,\lambda;x,y)$. This function is in principle a continuous real function, but becomes discrete once a given pixelization is assumed both on the detector plane and on the final datacube. For this reason, we are left with the problem of determining the function in its discretized approximation. This problem does not have a general analytical solution, because it involves computing the projection of datacube pixels onto detector pixels. Therefore, we approximate it in a Monte Carlo fashion. Specifically, per each pixel $(u_i,v_j,\lambda_k)$ on the image plane, we randomly draw photons inside the pixel in a uniform way, and then ray-trace onto the detector through the projection given by $\mathcal{G}$. In this way, we mimic a continuous distribution of photons (as is the case in real observations) despite the chosen pixelization. Once the photons have been projected onto the detector, we compute the fraction of photons which falls within the pixel $(x_m,y_n)$. This determines the coefficient $\mathcal{G}^\prime_\theta(u_i,v_j,\lambda_k;x_m,y_n)$ in the linear system of equations. Given the inherent stochasticity of this procedure, the determination of $\mathcal{G}^\prime$ will suffer from stochastic noise, which will depend on the number of photons used to sample the pixel. Therefore, in each problem defined by a given pixelization --- both in terms of number and of shape of the pixels --- one will need to determine $\mathcal{G}^\prime$ and study its numerical convergence. 

% *****************************************
\subsection{Tomographic solutions: filtered backprojection and iterative reconstruction}\label{sec:tomography_solution}

In the previous sections, we have presented the direct, intuitive solution of the problem based on computing the response function $\mathcal{G}$ and solving the system by matrix inversion. While conceptually simple, this poses  the following important issues at a computational level:
\begin{itemize}
\item the computation of $\mathcal{G}$ is not straightforward and requires evaluating how the photons of coordinates $(u,v,\lambda)$ get spread on the detector and to which pixel they contribute. Since the the projection mapping from the sky to the detector is not trivial, this must be performed numerically, by ray-tracing fictitious photons and compute in which pixel they will land. This can be performed using: (i) regular grid points, which however introduce artificial grid patterns, or (ii) randomly-sampling uniformly the photons one the sky, whose resulting Monte Carlo noise makes the inversion unstable. In either case, to overcome Poisson shot noise in the computation of $\mathcal{G}$, one must sample and ray-trace an extremely large number of photons, which becomes computationally very intensive; 
\item because the response matrix has size $N_{\rm unk}\times N_{\rm eq}$, where $N_{\rm unk}=U\times V \times l$ and $N_{\rm eq}$ is the number of linearly-independent equations, then the matrix size grow very quickly in size and the problem inversion becomes intractable even with a moderate datacube size of order $\mathcal{O}(10^2-10^3)^3=\mathcal{O}(10^6-10^9)$.   
\end{itemize}
In summary, the matrix that we seek to invert is often huge, ill-conditioned, and sometimes not even fully invertible.

To overcome these issues, we resort to typical reconstruction algorithms employed in tomography:
\begin{itemize}
    \item iterative reconstruction;
    \item filtered back-projection.
\end{itemize}

Both techniques --- that will be described more in detail in the following sections --- are based on the idea of forward-modelling the problem instead of inverting it. In other words, given our linear system $f=\mathcal{G}f^{\prime}$, instead of solving it via inversion $f^{\prime}=\mathcal{G}^{-1}f$, we seek to determine the approximated datacube $\hat{f}^{\prime}$ which minimizes the function $||f-\mathcal{G}\hat{f}^{\prime}||^2$, i.e. finding the datacube $\hat{f}^{\prime}$ whose forward-projection matches the observed data $f$. 

%\JZN{  I don't understand here.}\FSN{I will try to simplify the way it is written, but the idea is the following. Initially, we wanted to find the datacube $f^\prime$ by inverting the problem starting from the data (i.e. pixel flux values in the CCD images) $f$. Now instead we guess $f^\prime$ directly and forward-mode the problem transforming it  to $f$, compare with the data, and iterate until convergence is reached.}
This is done by introducing the normal equations $\mathcal{G}^\star\mathcal{G}f^{\prime}=\mathcal{G}^\star f$, where $\mathcal{G}^\star$ is the adjoint operator of $\mathcal{G}$\footnote{An adjoint operator $A^\star$ of an operator $A$ on a Hilbert space $\mathcal{H}$ is a unique operator that satisfies the relation: $\braket{Au,v}=\braket{u,A^\star v}$, for all $u,v\in \mathcal{H}$. It is the generalized `conjugate transpose' of $A$, often acting as a `backward' version of $A$ in functional analysis, quantum mechanics, and matrix theory.}  and $\mathcal{G}^\star f=\tilde{f}^\prime$ is a back-projection of the data onto a pseudo data-cube $\tilde{f}^\prime$, that differs from the true datacube $f^\prime$ by a convolution (see also Section~\ref{sec:filt_backproj}). Introducing the operator $\mathcal{G}^\star$ allows us to formulate the problem in a convenient way, as will be shown in Sections~\ref{sec:iterative_recon} and \ref{sec:filt_backproj}. 
In practice, since we do not need to store in memory the whole $\mathcal{G}$ matrix to perform its inversion, we implement the calculation of $\mathcal{G}$ and $\mathcal{G}^\star$ in a pixel-wise fashion by mapping the center of each voxel of coordinates $(u,v,\lambda)$ onto $(x,y)$ on the detector through the dispersion equations. While this relies on a regular grid, to alleviate potential artificial grid patterns and improve the smoothness of the resulting datacube by depositing the projection of the voxel onto the detector through a bilinear interpolation scheme, instead of a nearest grid point one. The possibility to perform the computation of $\mathcal{G}$ pixel by pixel without storing all the coefficients allows to significantly alleviate the computational needs of our machinery. In addition, we notice that forward-modelling the problem instead of inverting it allows to naturally treat the noise and the PSF in a consistent manner in the datacube and on the detector, which enables to drop the assumption of high SNR and be able to handle any type of sources (even pure noise).

% ***********************************************
\subsubsection{Iterative tomographic reconstruction}\label{sec:iterative_recon}

The first solution that we propose is a classical iterative problem. We start with an initial guess $\hat{f}^\prime_0$ and update it at iteration $i$ as $\hat{f}^\prime_{i}=\hat{f}^\prime_{i-1}+\Delta \hat{f}^\prime$, where the correction $\Delta \hat{f}^\prime$ is computed based on the residuals $r=f-\mathcal{G}\hat{f}^\prime$. In particular, the simplest minimization algorithm adopts a classical gradient descent approach, in which the function $||f-\mathcal{G}\hat{f}^\prime||^2$ is minimized by following the direction of the gradient $\nabla=\mathcal{G}^\star (\mathcal{G}\hat{f}^\prime-f)$.\footnote{The gradient can be derived as follows. Let us first express the target function to be minimized as: $||f-\ \mathcal{G}f^\prime||^2=(f-\mathcal{G}f^\prime)^\star~(f-\mathcal{G}f^\prime)=f^2 -2f^\star\mathcal{G}f^\prime + f^\star\mathcal{G}^\star\mathcal{G}f$. Let us now consider the following identities: (i) $\nabla_f(f^\star\mathcal{G}f)=2\mathcal{G}f$, and (ii) $\nabla_f(\mathcal{G}^\star f)=\mathcal{G}$. Taking gradients w.r.t. to $f'$ and substituting the two identities above gives the final result.}
%\JZN{  why this is the gradient?} \FSN{The gradient comes from: $||f-\ \mathcal{G}f^\prime||^2=(f-\mathcal{G}f^\prime)^\star~(f-\mathcal{G}f^\prime)=f^2 -2f^\star\mathcal{G}f^\prime + f^\star\mathcal{G}^\star\mathcal{G}f$. Let us now consider the following identities: (i) $\nabla_f(f^\star\mathcal{G}f)=2\mathcal{G}f$, and (ii) $\nabla_f(c^\star f)=c$. Taking gradients w.r.t. to $f'$ and substituting the two identities above gives the final result.}
%\JZN{Is $f$ a convex function?I don't know, is it possible to always have one global minimum? I guess so in most of the physical problems?}\FSN{It does not need to be strictly convex, and it will likely have multiple local minima, which potentially complicate the problem. However, there exist numerical techniques that go and look for the global minimum even in a case in which the parameters space has a complex topology.}
In this way, the update formula introduced above reads: $\hat{f}^\prime_{i}=\hat{f}^\prime_{i-1}+\alpha \nabla$, with $\alpha$ a parameter controlling the step size taken at each iteration. More efficiently, one can adopt either least-square minimization algorithms based on conjugate gradients --- that explore optimal directions in the solution space and feature a faster convergence --- or the Lucy-Richardson reconstruction \citep{Richardson1972,Lucy1974} using likelihood maximum estimation under the assumption of Poisson noise. Specifically, we resort to the latter due to its efficiency and stability, and because it incorporates the correct Poisson model for photon counts. %\FSN{Add here: initial guess (1 everywhere and number of iterations needed. Tested so far: 100 iterations, takes less than 10 minutes and the reconstruction is not bad.}

% ***********************************************
\subsubsection{Filtered back-projection}\label{sec:filt_backproj}

The second solution proposed herein consists of filtered back-projection \citep[see e.g.,][and references therein]{Beckmann2024}. In the first step, one computes a pseudo-datacube $\tilde{f}^\prime=\mathcal{G}^\star f$ by back-projecting $f$ through the adjoint operator $\mathcal{G}^\star$. However, this operation has a blurring effect on the pseudo-datacube, as $\mathcal{G}^\star f= \mathcal{G}^\star (\mathcal{G}f^\prime)=(\mathcal{G}^\star \mathcal{G})f^\prime$, and $\mathcal{G}^\star \mathcal{G}$ acts like a convolution. Because in Fourier space $\mathcal{F}(\mathcal{G}^\star \mathcal{G})\propto 1/|k|$ (where $\mathcal{F}$ denotes `Fourier Transform'), to undo the blurring effect one Fourier-transform the pseudo-datacube, multiplies it by $|k|$ and then inverse-transform it back to real space: $f^\prime=\mathcal{F}^{-1}\left(|k|\mathcal{F}(\tilde{f}^\prime)\right)$. 

%\JZN{  I am lost here.} \FSN{You try to solve the problem in Fourier space rather than in real space, as in the previous case. Doing it in Fourier space introduces a spurious effect, that you have to deconvolve. We actually don't use this technique here (I tested it but doesn't work as well as the other), so we could move this paragraph to an Appendix.}

While filtered back-projection is efficient and solves the problem in just one step, it assumes ideal conditions, that may not be suited to our spectrograph cases, which has finite detector pixels, noise, wavelength-dependent shifts, among others. On the other hand, iterative methods can be more expensive as they require many iterations, but can naturally incorporate numerical tricks that stabilize the solution, such as enforcing flux positivity and apply regularization. In addition, they can deal with the noise statistics and forward-model other effects, such as the PFS and field distortions. 

For these reasons, we benchmark in this paper only iterative reconstructions, and leave filtered back-projection as an alternative to be explored in future works.

% *****************************************

\subsection{Numerical experiments of iterative reconstructions}\label{sec:exp}

As a proof of concept, we performed a few numerical experiments, and discuss in this section the results. The experiments are set up as follows:
\begin{itemize}
    \item given the problem setup and once established the pixelization of the reconstructed image, we compute the number of rotations needed to fully specify the problem at a mathematical level, following Eq. \ref{eq:num_rot};
    \item we initialize a collection of source photons on the sky plane (i.e., of coordinates $(u,v,\lambda)$) according to a given sky model;
    \item we project the photons onto the detector by simulating the spectrograph dispersion as described in Section~\ref{sec:spec_dispersion}. We repeat this procedure for the different orientations of the spectrograph, producing a number of detector images equal to the number of rotations;
    \item we perform the datacube reconstruction, given the input set of detector images. As discussed previously, this can be done either by simulating the spectrograph spreading function $\mathcal{G}$ (see Section~\ref{sec:spec_spread_function}) and by inverting the system of questions (Eq.~\ref{eq:main_system}) --- at high computational cost --- or by performing tomographic reconstruction. Herein, we test only the latter.
\end{itemize}

%We perform a variety of numerical experiment, that we describe in what follows. 
%The main difference between the distinct experiments lies in the setup---both in terms of pixelization of the image and the CCD, as well as of the photon sources --- while the practical procedure remains the same. 
In these tests, we express all the physical quantities --- both $(u,v,\lambda)$ coordinates and flux --- in arbitrary units, often normalized in such a way that the maximum equals one for simplicity.

\begin{figure*}
    \centering
    \includegraphics[width=0.85\linewidth]{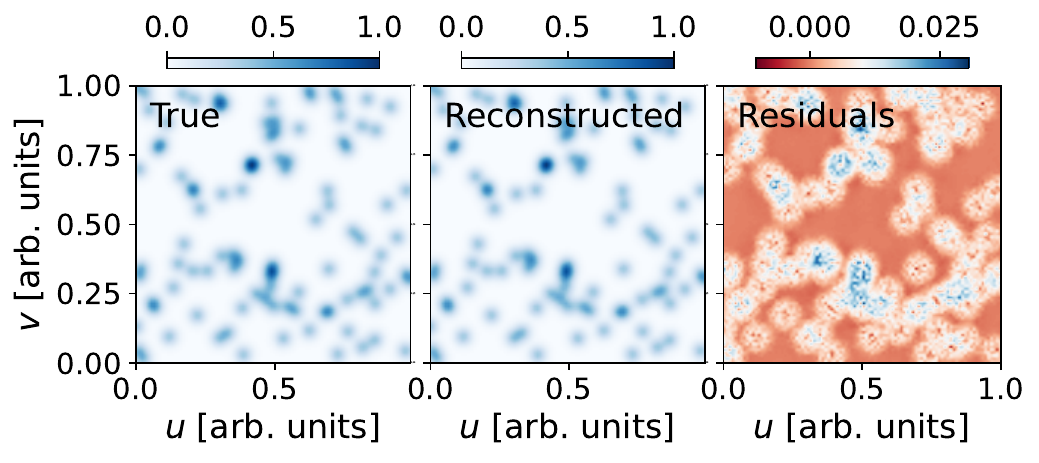}
    \includegraphics[width=0.85\linewidth]{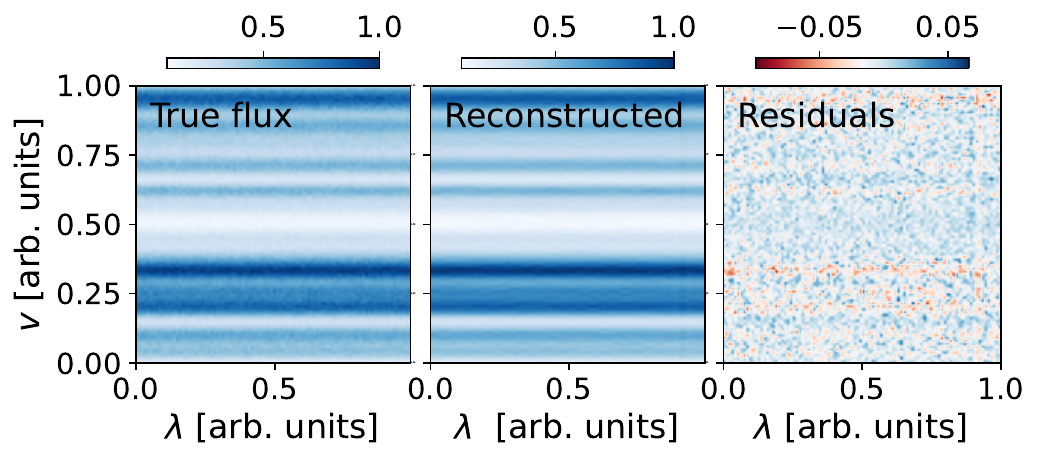}
    \caption{Reconstructed images of a full datacube containing $100$ stars with uniform spectra, for the case in which the dispersion element is rotated with respect to the detector. Top: projection of the datacube on the $(u,v)$ plane, for the true datacube (left), the reconstructed datacube (center), and their resisuals (right). Bottom: same as top, but for the $(v,\lambda)$ plane}
    \label{fig:100stars}
\end{figure*}
%\begin{figure*}
%    \centering
%    \includegraphics[width=\linewidth]{results_radec_100_stars_unif_buf0.2_fieldrot.pdf}
%    \caption{Caption}
%    \label{fig:100stars_radec_fieldrot}
%\end{figure*}

\begin{figure*}
    \centering
    \includegraphics[width=0.85\linewidth]{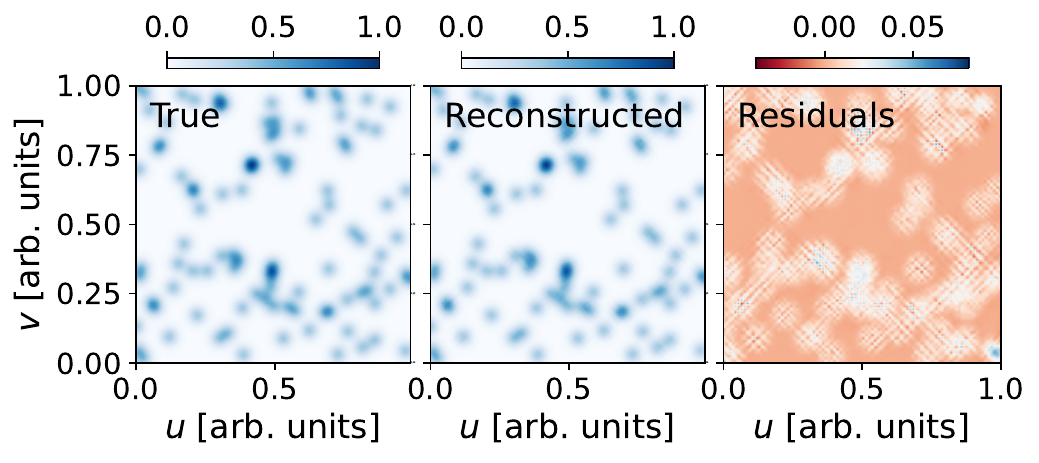}
    \includegraphics[width=0.85\linewidth]{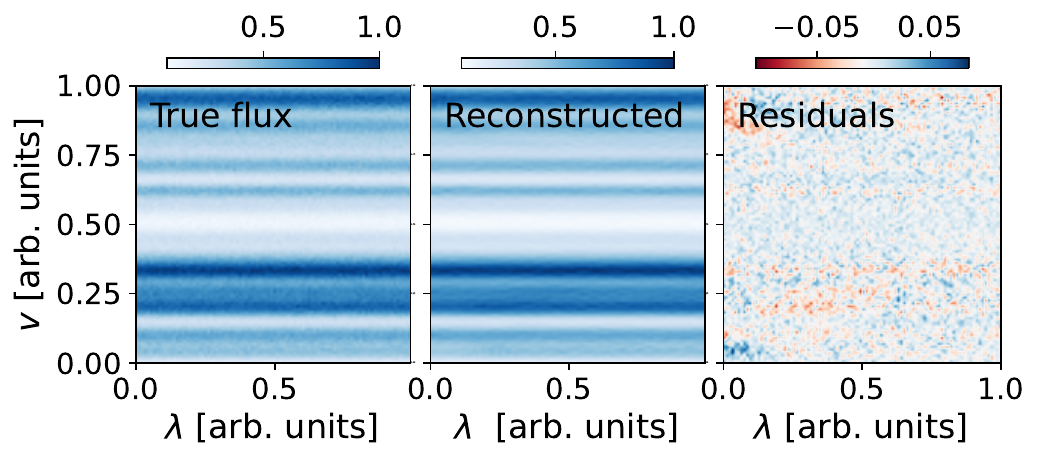}
    \caption{Same as Figure~\ref{fig:100stars}, but for the case in which the aperture is rotated with respect to the field}
    \label{fig:100stars_fieldrot}
\end{figure*}
%\begin{figure*}
%    \centering
%    \includegraphics[width=\linewidth]{results_ralam_100_stars_unif_buf0.2_fieldrot.pdf}
%    \caption{Caption}
%\label{fig:100stars_declam_fieldrot}
%\end{figure*}

\begin{figure*}
    \centering
    \includegraphics[width=0.85\linewidth]{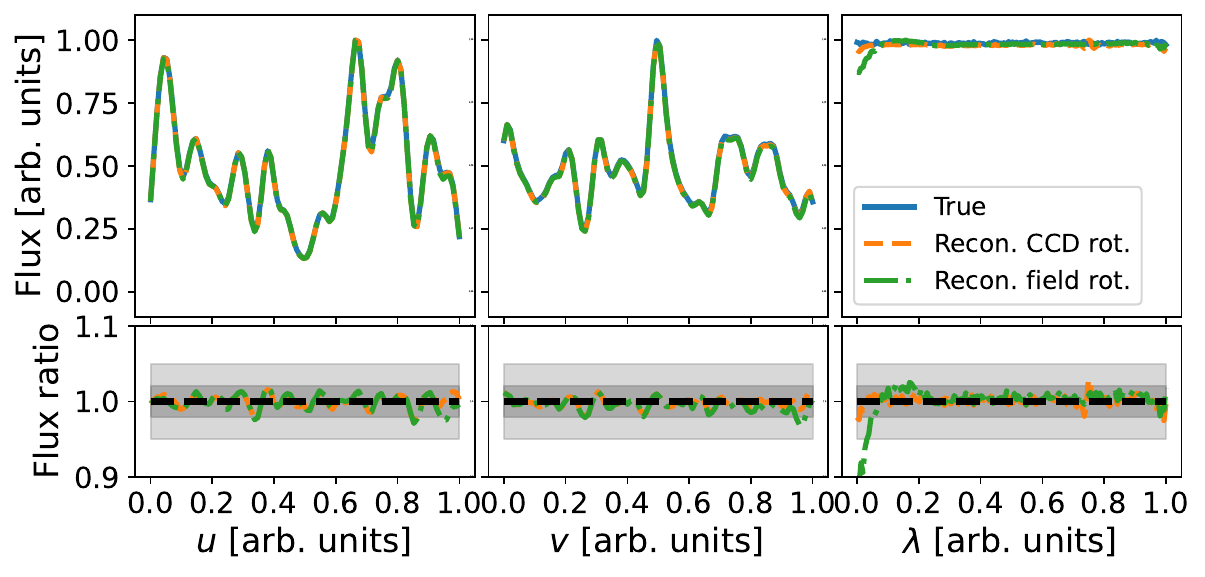}
    \caption{Reconstructed flux profiles of a full datacube containing $100$ stars with uniform spectra, along the $u$ (left), $v$ (center), and $\lambda$ (right) direction. We display the true profiles as a blue solid line, the reconstructed profile obtained by rotating the dispersion element with respect o the detector as an orange dashed line, and the reconstructed profile obtained by rotating the aperture wit respect to the field as a green dotted-dashed line. The top panels show the actual flux profiles, while the bottom panel show the ratios between the reconstructed and the true curves.}
    \label{fig:100stars_ratios}
\end{figure*}

%

%***************************************
%\subsubsection{$100$ stars with uniform spectra}

We inject $100$ stars in the field, each one with a uniform spectrum. This configuration allows us to test a realistic case in which we target a crowded field with contaminated spectra filling the whole detector. The latter aspect is especially important, as it allows us to study the edge effects in the reconstruction procedure. We model each star on the sky as a collection of $n=10^8$ photons, with $(u,v)$ coordinates drawn from a $2$D Gaussian distribution which mimics the PSF, while we perform a uniform random sampling along the spectral direction. We do not inject noise, whose effect will be studied in future work. We assume a datacube of size $(U,V,l)=(160,160,160)$ and a detector of size $(X,Y)=(160,160)$ pix, which implies $n_{\rm rot}=160$. In this case we test reconstruction based on either rotating the telescope or rotating the dispersion element. The first reconstruction attempt that we performed results in qualitatively correct reconstructed datacubes, with stars located in the correct positions and uniform spectra, but with clear numerical artifacts. These arise from the fact that a large portion of the spectra of the stars towards the edges of the FoV fall outside the detector. While from the mathematical perspective offered by the linear system of Eq.~\ref{eq:main_system} it is sufficient that the information from each datacube pixel appears at least once to correctly solve the problem, at a numerical level the algorithm struggles to converge well to the correct solution in all the pixels, and numerical spurious effects appear. This aspect consists of one important limitation of the novel technique introduced herein. While we will discuss this issue in detail in future work, here we test the approach of restricting the mock observations to field which fills only the central part of the detector, leaving an empty buffering zone in the outer part to avoid edge effects. We conservatively assume a $20\%$ buffering zone at each boundary. For the case in which we rotate the spectrograph with respect to the detector, the upper and lower panels of Figure~\ref{fig:100stars} show the projection onto the $(u,v)$ plane and analogous projections on the $(v,\lambda)$ plane of the true (left) and reconstructed (mid) datacubes, as well as the residuals (right) at the end of the reconstruction.  Figure~\ref{fig:100stars_fieldrot} shows the same, but this time rotating the whole telescope. 

We report an excellent convergence of the reconstruction procedure, with very tiny residuals. Figure~\ref{fig:100stars_ratios} displays the projections of the true (blue solid) and reconstructed (orange dashed and green dotted-dashed for rotations of the spectrograph or of the telescope, respectively) datacubes along the $u$ (top left), $v$ (top mid), and $\lambda$ (top right) coordinates, together with their ratios (reconstructed/true) in the bottom panels. The achieved accuracy is excellent, with deviations generally within $\sim 2\%$ (darker gray dashed area) in the central region all over the field and the wavelength range. While the assumed buffering zone is quite large and reduced the effective used detector area to just $\sim 36\%$ of the original, in future work we will investigate numerical techniques aiming at improving the reconstruction and relax the assumption made here, trying to get rid of or at least reduce the buffering zone.   

The $500$-iteration reconstruction presented here takes $<0.5$\,hours on a normal MacBook Pro M2 laptop, with 10 cores and 16GB RAM. This shows that applying iterative reconstruction is computationally feasible even in ordinary devices.

% ***************************************

%\subsection{Noise model}\label{sec:noise}
%\JZN{going to read some literature}

%--------------------------------------------------------------------
\section{Discussion}
\label{sec:discussion}

In this section, we discuss different aspects related to the scientific exploitation and limitations of \textsc{rossini}.

\subsection{A cheap and efficient all-in-one instrument}

We argue that \textsc{rossini} is a powerful and cheap all-in-one-instrument: it can observe in (i) imaging mode, by not inserting the dispersion element, (ii) slitless spectroscopy mode, keeping fixed the rotation angle of the dispersion element, and (iii) IFS mode, performing observations at various rotation angles and applying a reconstruction technique to generate the datacube. As such, it allows the observer to have different types of observations with just very simple configuration changes. Compared to state-of-the-art IFS technologies, it therefore represents a cheap alternative, that utilizes existing slitless spectrographs and do not need fragile and expensive IFUs, at the only cost of mounting the dispersion element on a rotating devise and/or having a field rotator.  %Here, we argue that a simple modification of the mechanical design can add the 

%For comparison, the cost of the SIWFT spectrograph installed on the Hale Telescope at the Palomar Observatory is estaimted to be of order few million dollars\footnote{\url{https://noirlab.edu/science/resources/legacy/swift/costs}}.

\subsection{Science cases}

In what follows, we describe a number of science cases that would optimally benefit from this new spectrograph design, beyond the efficiency and cost arguments:
\begin{itemize}
    \item {\bf faint object spectroscopy:} Just like other slitless spectroscopy, \textsc{rossini} does not inherit any fluxloss caused by slit or fiber. In particular, the simple design of \textsc{rossini} has the shortest light path without passing extra optical elements like image slicer. It can potentially deliver a higher efficiency and throughput which is important for faint targets like ultracool dwarfs and high-redshift galaxies. Moreover, the sky background will be automatically recorded on the sky area in the datacube, and can be averaged in a large area and then subtracted from the object spectra without introducing dithering overheads or extra sky exposures.
    \item {\bf dense fields:} globular clusters or galaxy clusters have inhomogeneous source distributions. \textsc{rossini} can be optimally exploited to obtain a datacube with a higher spatial resolution towards the high density locations as compared to standard IFU, as well as optimally alleviate the spectra contamination that would arise in this case from standard slitless spectroscopy;
    \item {\bf extended objects:} IFS has become a standard technique to perform spatially-resolved imaging and spectroscopy of extended sources such as nearby galaxies and nebulae. \textsc{rossini} can be naturally used to optimize the observing strategy, by choosing a proper adaptative pixelization, similarly to what discussed above for globular clusters. For instance, one may be have greater interest in resolving well spatial or spectral features---e.g. star-forming regions or the central part where the influence of the supermassive black hole is larger on the sky plane, the Balmer break and emission/absorption lines in the spectral direction---while having lower resolution in low-density patches of the galaxy or in slowly varying regions of the spectrum.
    %\item {\bf galaxy clusters:} as for globular clusters, galaxy cluster have higher density of galaxies and gas towards the center, therefore the same arguments presented above apply here as well;    
    \item {\bf cosmology and galaxy surveys}: state-of-the-art cosmological large-scale structure surveys, such as DESI and \textit{Euclid} are being performed using fibers (DESI) or grism spectroscopy (\textit{Euclid}). In this field, \textsc{rossini} has the potential to improve over as: (i) it can naturally perform spectra decontamination, as was already proposed in the original \textit{Euclid} design (ii) it does not need previous target selection, (iii) if previous target selection based on photometric surveys is available, it could offer a way forward to solve the `fiber collision' issue, consisting in the failure in observing two objects which are separated on the sky by a distance which is smaller than the physical separation of two fibers, and therefore only one of the two can be actually observed. Beyond galaxy surveys, \textsc{rossini} can be employed to generate datacube in line intensity mapping experiments. Also in this case, if an acquisition is available beforehand, it can be used to optimally tailor the pixeling strategy in such a way to target denser regions of the large scale structure (e.g., clusters and filaments) with higher spatial and spectral resolutions, and low-density environments (e.g. voids) with lower resolution. 
    \item {\bf a full-sky IFS survey:} Not pointing only at targets of interest, having an IFS survey for the entire sky until a certain brightness level would contribute to the legacy value for the whole astronomical society. Considering the extremely small FoV and high cost of traditional IFSs, \textsc{rossini} offers an opportunity to achieve this goal in a time scale of few years, depending on the required magnitude limit, resolution, and numbers of facilities installed with \textsc{rossini}.

\end{itemize}

%\FSN{I like it very much the way you have re-ordered this section.}

%high S/N situation: solar, bright nebular, globular clusters, fiber collision and intensity mapping (cosmology and LSS).

%\FSN{Just throwing all the ideas in, needs polishing and reordering.}
%Discuss limitations, efficiency, costs as compared to real IFU. Highlight the potential all instruments in one by having a simple slitless if no rotation, IFS if rotation. 

\subsection{Deployment on future telescopes}

Among the strengths of the rotational slitless spectroscopy presented in this work is the fact that it does not require any particular technological development, except for the implementation of a rotating device that allows the user to rotate the dispersion element. In this sense, the real engineering challenge will consist in controlling the rotation angle with high precision, in such a way to guarantee the largest possible number of rotations, and hence, of angular and spectral resolution. 

As already anticipated, both in space-based and ground-based telescopes, a natural way of applying this technique would consist in performing different exposures at different telescope rotation angles with respect to the sky, by rotating the spacecraft around the roll angle or using the field rotator in a ground-based telescope. Therefore, this poses no issues with existing technology.

In addition, in space-based telescopes (\textit{HST, JWST, Euclid}) the possibility of performing spectra decontamination through different dispersion directions is controlled by different grism mounted on a wheel which allows to move change the grism in an efficient fashion. Moving forward from this idea, one will need to replace the grisms wheel with a motorized device which allows a single grism. Compared to e.g. the state-of-the-art IFU NIRSpec instrument mounted on {\it JWST}, \textsc{rossini} will offer the advantage of being able to survey a much wider field and be less expensive and delicate. In ground-based telescopes, no such idea has been explicitly deployed, as the spectra decontamination is achieved by rotating the whole instrument, as mentioned above. Therefore, also in this case, a novel device allowing the observer to change the dispersion direction on the detector will needed to de developed. We also notice that \textsc{rossini} can naturally couple with adaptive optics.

% ****************************
\subsection{Combining spectrograph and telescope rotations}

%\JZN{I feel this combination is unnecessary.} \FSN{Practically speaking, I feel like only the field rotation will be used, as it can be done right away if you have  a field rotator, without the need of building a new instrument. However, imagine a scenario in which you want to observed a very large FoV, with very high spatial and spectral resolution, for which you need 10,000 rotations. This would imply, an angular precision $\Delta\theta=360^\circ/10,000=0.036^\circ\sim 2.16^\prime$. This may or may not be technically challenging. If instead you decompose $10,000=100\times100$, you can do just $100$ rotations of each type and the precision requirement is much lower.}

While in the previous sections we discussed separately the cases of rotating the spectrograph with respect to the detector or the whole telescope, we notice that the two techniques can naturally be combined. In fact, they provide fully independent ways of constructing detector images, and therefore can be used in synergy. This makes it possible to:
\begin{itemize}
    \item effectively increase the number of effective of independent images, thereby reaching arbitrarily high resolution and overcoming potential mechanical limitations, such as the precision of precision controlling the rotations; 
    \item alleviate potential systematics from each of the two methods, such as e.g. the limited mixing of the information of detector pixels close to the rotation center;
    \item reduce the persistence effect on the detector caused by stationary bright objects, especially for infrared detectors.
    %\item given the maximum number of rotations $n_{\rm rot,max}$ allowed by the mechanical devices, have a total number of images of $n_{\rm rot,max}^2$, which allows the user to achieve a much higher spatial and spectral resolution;
    %\item once the resolution is fixed, the same number of independent images can be achieved through a combination of the two rotation movements, lowering the number of rotations in each single degree of freedom, and hence also alleviating the precision requirements in the rotations.
\end{itemize}

% ****************************
\subsection{Initial guess for the reconstruction and flux calibration}
The iterative reconstruction procedure adopted herein needs initializing the datacube to some initial guess. In this sense, we straightforwadly initialize the datacubes to values equal one everywhere, but we have explored different initial guesses and found that the convergence is generally robust against the choice. This leaves us with the problem of the arbitrarity of the normalization of the final datacubes. In our simulations, we can manually enforce the right normalization, given that we know the ground truth. In real observations, however, we will clearly not be in this situation. Nonetheless, as is done in any standard astronomical observation, we can rely on standard stars with known magnitudes and fluxes to convert the arbitrary voxel values into physically meaningful quantities and correctly renormalize our cube.
Additionally, we can exploit the fact that \textsc{rossini} is naturally also an imager, to acquire one or few images at selected wavelengths, and use them to have an initial guess which is closer to the final answer, as well as have a first relatively correct normalization.

\subsection{Limitations}

The main conceptual limitation associated with \textsc{rossini} is the accuracy of the datacube reconstruction. In particular, while we have tested here just a few idealized cases, many additional effects need to be taken into account.
On the technological side, we notice that rotating the grism potentially introduces calibration, stability, and optical aberrations issues, that will be needed to be thoroughly addressed. %Furthermore, as stated in the previous Section, the precision of the rotating device will effectively dictate the number of possible rotations, and therefore, the maximum number of spaxels allowed in the reconstructed datacube. 

\subsection{Future work}

As anticipated, this paper presents just a proof of concept of this novel rotational slitless spectrograph. As such, a considerable amount of work is left ahead, to be done in the near future. In particular, we foresee the following development items:
\begin{itemize}
    \item testing the reconstruction techniques in more realistic cases, including noise, an explicit modelling of the PSF, and other instrumental effects. In particular, we plan to assess the accuracy of the reconstruction on real datacubes, such as e.g. the ones from the Manga survey;
    \item testing the reconstruction technique with adaptative pixeling --- mathematically described in Section~\ref{sec:concept} --- and address its practical feasibility; 
    \item explore different reconstruction techniques beyond the simple Lucy-Richardson reconstruction tested herein: more advanced optimization schemes can be adopted, both deterministic relying on maximum-likelihood estimation, and Bayesian based on Markov Chain Monte Carlo algorithms. In particular, regarding the latter, by treating the flux in each spaxel as a variable one can deploy efficient Monte Carlo techniques such as Hamiltonian Monte Carlo and No U-Turn Sampling, and find the full posterior distributions. This allows one to retrieve not only the a single flux values for each spaxel, but also the associated uncertainty stemming from the reconstruction. In addition, the intrinsic quality of the reconstruction can be improved by means of filtering and convolution/deconvolution techniques, which help the convergence in regulating the sharpness of the reconstructed structure and alleviate aliasing effects; %\JZN{more explain for me here please} \FSN{Which part? HMC/NUTS? Or deconvolution?}
    \item assess the best strategy to handle edge effects, beyond conservatively assuming a buffering zone on the detector;
    \item deploy a prototype of this instrument and the related reconstruction, and showcase the practical on-sky feasibility of this technology.
\end{itemize}
 
%-----------------------------------------------------------------

\section{Conclusions}
\label{sec:conclusion}

In this work, we presented a novel concept of \textsc{rossini}: the ROtational Slitless Spectrograph for INtegral field spectroscopy and Imager, combining slitless spectroscopy with a rotation of the field or the dispersion direction and tomographic reconstruction, which allows us to perform IFS without explicitly having IFUs. In particular, we exploit the fact that by rotating the whole telescope about the pointing axis, or changing the dispersion direction on the detector through a rotation of the optical element (e.g., a grism), we obtain different, statistically independent detector images. We showed that, assuming a pixelization of the sky volume that we want to produce a datacube of, and with a sufficient number of rotations dictated by the relative size of the datacubes, it is possibly to mathematically express the flux in each datacube voxel as linear combination of the flux observed in the distinct pixels on the detector, and that therefore the problem reduces to solving a linear system of equation. From another perspective, we argue that reconstructing the datacube by means of this novel spectrographic device falls within the class of `tomographic reconstruction' methods, in which $2$D projections of a $3$D volume are used to reconstruct the latter. In particular, the slitless rotational spectrograph is conceptually equivalent to medical CT. 

To showcase the potential of this novel device, we test the reconstruction procedure by simulating
%on two toy model cases: a field with only one star and a Gaussian-shaped spectrum, 
%\JZN{remove the one star case, if you accept my suggestion}\FSN{Fully agree} 
a crowded field with $100$ stars and uniform spectra. We first simulate the detector images, and then apply an iterative reconstruction method explicitly tailored to Poisson photon counts statistics. In both cases, we managed to reconstruct the input datacube with excellent accuracy ($2-5\%$ maximum deviations), using only $500$ iterations and no further refinements. To prevent numerical artifacts stemming from edge effects and the consequent loss of information due to large portions of spectra falling outside the detector, we show that assuming a $20\%$ buffering zone in the outer part of the detector is sufficient to achieve high accuracy.

We stress that this paper presents only a proof of concept for this novel instrument and already achieves excellent accuracy adopting simple reconstruction algorithms, and that future work will aim at significantly improving it. 

The \textsc{rossini} method presented herein promises to provide a cheap alternative to classical IFS, and represents an interesting way forward to efficiently survey large FoVs. 

% ************************************
% ************************************
\begin{acknowledgements}
JYZ acknowledges support from the Western Postdoctoral Fellowship provided by Western University. JYZ also acknowledges support from the Agencia Estatal de Investigación del Ministerio de Ciencia, Innovación y Universidades under grants PID2019-109522GB-C53, PID2022-137241NB-C41, as well as from the European Union ERC AdG SUBSTELLAR grant agreement number 101054354. FS acknowledges the support from the Swiss National Science Foundation (SNSF) 200021\_214990/1 grant and from the IFPU Fellowship scheme. The authors are grateful to the Adolfo Suárez Madrid-Barajas Airport, where the idea was conceived, and to the cafeteria patio at the Instituto de Astrofísica de Canarias, where fruitful discussions on this project took place.
\end{acknowledgements}

%\section{Appendixes}

%\nocite{*}

\bibliography{main}% Produces the bibliography via BibTeX.

\end{document}